\documentclass[graybox]{svmult}

\usepackage{mathptmx}       
\usepackage{helvet}         
\usepackage{courier}        
\usepackage{type1cm}        
\usepackage{makeidx}         
\usepackage{graphicx}        
\usepackage{multicol}        
\usepackage[bottom]{footmisc}

\newcommand{\kms}{\mbox{km s$^{-1}$}}

\newcommand{\arcsec}{\mbox{$''$}}

\newcommand{\arcmin}{\mbox{$'$}}

\newcommand{\degree}{\rm^{o}}
\newcommand\apj{{ApJ}}%
\newcommand\apjl{{ApJ}}%
\newcommand\apjs{{ApJS}}%
\newcommand\aap{{A\&A}}%
\newcommand\nat{{Nature}}%
\makeindex             

\begin{document}

\title*{Molecular gas properties of galaxies: The SMA CO(2-1) B0DEGA legacy project}
\author{D. Espada, S. Martin, P.-Y. Hsieh, P. T. P. Ho, S. Matsushita, L. Verdes-Montenegro, J. Sabater, S. Verley, M. Krips, V. Espigares}
\authorrunning{D. Espada et al.}

\institute{D. Espada \at Instituto de Astrof{\'i}sica de Andaluc{\'i}a - CSIC, Apdo. 3004, 18080 Granada, Spain. \email{daniel@iaa.es}
}
%
%
\maketitle



\abstract*{
In the last two decades high resolution ($<$ 5\arcsec) CO observations for $\sim$ 150 galaxies have provided 
a wealth of information about the molecular gas morphologies in the circumnuclear regions. 
While in samples of 'normal' galaxies the molecular gas does not seem to peak toward the nuclear regions for about 50\% of the galaxies, barred galaxies and mergers show larger concentrations. However, we do not exactly know from an observational point of view how the molecular gas properties of a galaxy evolve as a result of an interaction.
Here we present the SMA CO(2--1) B0DEGA (Below 0 DEgree GAlaxies) legacy project in which we are imaging the CO(2--1) line of the circumnuclear regions (1\arcmin) of a large ($\sim$ 70) sample of nearby IR-bright spiral galaxies, likely interacting, and that still remained unexplored due to its location in the southern hemisphere. We find different molecular gas morphologies, such as rings, nuclear arms, nuclear bars and asymmetries. We find a centrally peaked concentration in about 85\% of the galaxies with typical size scales of about 0.5 -- 1~kpc. This might be related to perturbations produced by recent interactions. 
}

\abstract{
In the last two decades high resolution ($<$ 5\arcsec) CO observations for $\sim$ 150 galaxies have provided 
a wealth of information about the molecular gas morphologies in the circumnuclear regions. 
While in samples of 'normal' galaxies the molecular gas does not seem to peak toward the nuclear regions for about 50\% of the galaxies, barred galaxies and mergers show larger concentrations. However, we do not exactly know from an observational point of view how the molecular gas properties of a galaxy evolve as a result of an interaction.
Here we present the SMA CO(2--1) B0DEGA (Below 0 DEgree GAlaxies) legacy project in which we are imaging the CO(2--1) line of the circumnuclear regions (1\arcmin) of a large ($\sim$ 70) sample of nearby IR-bright spiral galaxies, likely interacting, and that still remained unexplored due to its location in the southern hemisphere. We find different molecular gas morphologies, such as rings, nuclear arms, nuclear bars and asymmetries. We find a centrally peaked concentration in about 85\% of the galaxies with typical size scales of about 0.5 -- 1~kpc. This might be related to perturbations produced by recent interactions. }

\section{Introduction}
\label{sec:1}

The molecular gas is one of the dominant components of the interstellar medium (ISM) in the inner few kiloparsecs of spiral galaxies.
It is not only an essential ingredient to understand the structure and kinematics of the ISM there, but it also constitutes the fuel for future star formation (SF). 
Molecular gas, due to its dissipative nature, efficiently loses angular momentum via gravitational torques and may fall into the central hundreds of parsec. This gaseous component is eventually one of the main drivers for the existence of a central starburst and/or the fuelling an active galactic nuclei (AGN)  \cite{1989Natur.338...45S}.  As a result, the concentration index in barred galaxies is seen to be larger than in non-barred galaxies \cite{1999ApJ...525..691S,2005ApJ...632..217S}. Galaxy-galaxy interactions are expected to play a major role in triggering the formation of bars, resulting in larger central gas concentrations in each individual galaxy. Then, the merging of two galaxies can lead to a violent starburst episode \cite{1991ApJ...370L..65B,2008ApJS..178..189W}.

We aim to study the circumnuclear morphologies of the bulk of the molecular gas and its relation with both intrinsic (i.e stellar morphology) and external variables (i.e. environment). 
High resolution molecular gas studies are needed.  However, the limited number of observed galaxies to date ($\sim$150) and large number of variables at play seems insufficient to understand how they affect the general properties of the bulk of the molecular gas.  The most numerous high resolution CO studies (best tracer at our disposal to trace the bulk of the molecular gas) of nearby spiral galaxies performed to date are: NRT-OVRO  (N=20 galaxies, \cite{1999ApJ...525..691S}), BIMA-SONG (N=44, \cite{2003ApJS..145..259H}), and the NUGA projects  (N=28, \cite{2003A&A...407..485G} and references therein). 
These images reveal a wide variety of morphologies in the circumnuclear molecular disks. 
BIMA-SONG is one of the largest compilation of CO interferometric data to date for 'normal' galaxies, and show that molecular gas is not centrally peaked in 55\% of the galaxies \cite{2003ApJS..145..259H}. 
However, this result is not likely valid for other samples of galaxies with other environmental properties.
We aim to enlarge the number of observed galaxies that are likely suffering recent interactions (although not mergers), which in addition to previously studied samples, will allow us to study how interactions modify the molecular gas concentration.

\section{The B0DEGA Sample}
\label{sec:sample}

We focus on a sample of: $i)$ nearby galaxies, with recession velocities $V$ $<$ 7000 \kms, being the bulk of galaxies at about 2000~\kms ;  $ii)$ IR-bright, following the criterion: 2.58 $\times$ S$_{60\mu m}$ + S$_{100\mu m}$  $>$ 31.5~Jy ($S_{60\mu m}$ and $S_{100\mu m}$ are the IRAS flux densities in the IRAS Point Source Catalog \cite{1988iras....7.....H}) so that CO(2--1) fluxes are high enough to be easily detected with the Submillimeter Array (SMA) in a reasonable amount of time of about 2 -- 3~hours and with resolutions better than $\sim$ 5\arcsec;  
 $iii)$ located in the southern hemisphere  and observable from SMA (45$\degree$ $<$ $\delta$ $<$ 0$\degree$). 
 We use the latter criterion because the lack of mm/submm interferometers in the southern hemisphere has prevented to study detailed molecular gas properties of many southern galaxies. 
The SMA can observe sources down to $\delta$ = -45$\rm ^o$, and it is revealing the molecular gas properties of many unique southern sources (for example Centaurus A, \cite{2009ApJ...695..116E}). 

With these criteria it yields a total of N = 134 galaxies, out of which only 14 had high-resolution CO(2--1) maps in the literature. Here we present data for a subsample of the IR brightest $N$ = 30 B0DEGA galaxies. Further observations for other galaxies in the sample are being carried out at the SMA as a legacy project.

\begin{figure}[]
\begin{center}
\includegraphics[width=8cm,angle=-90]{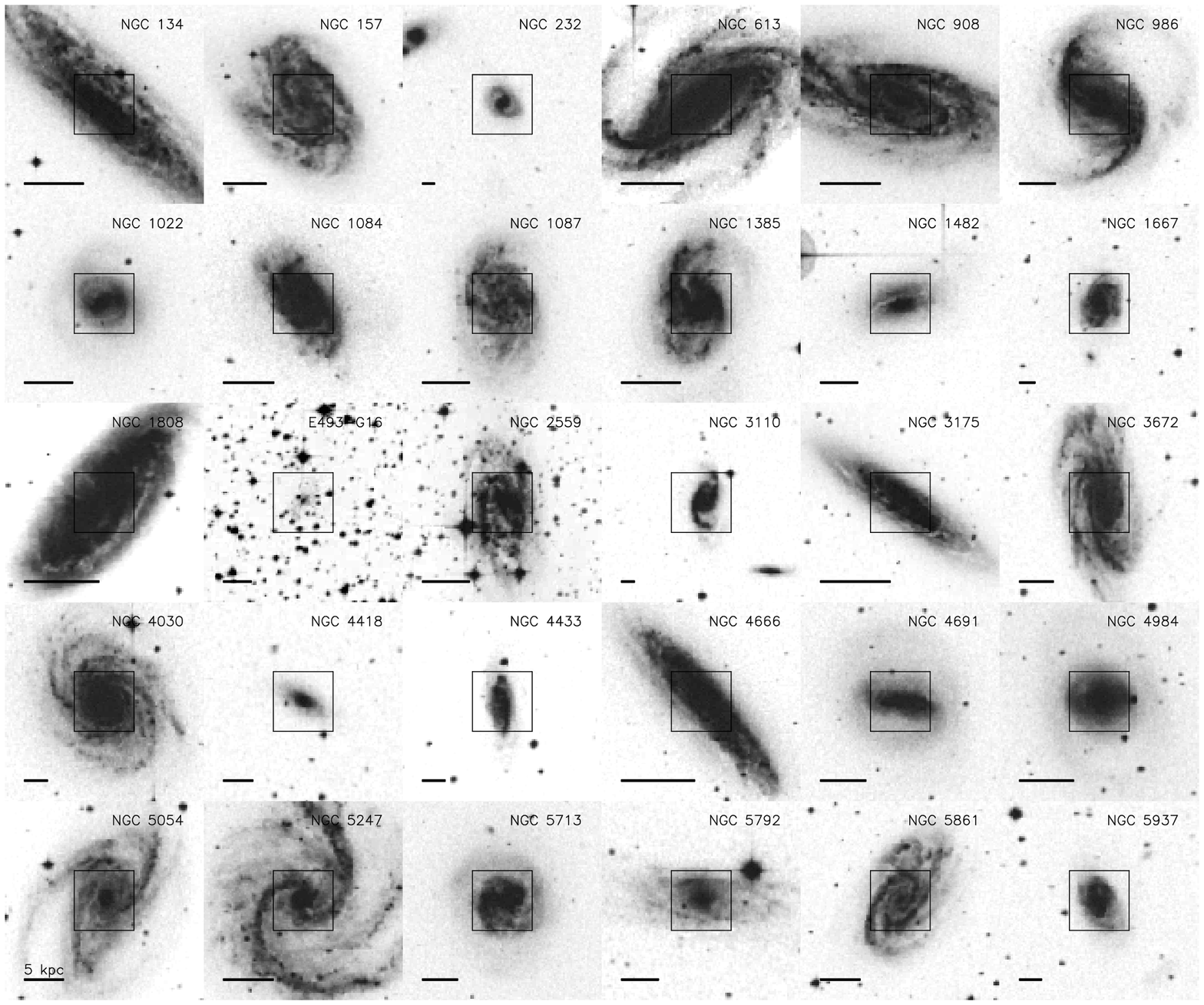}
\includegraphics[width=8cm,angle=-90]{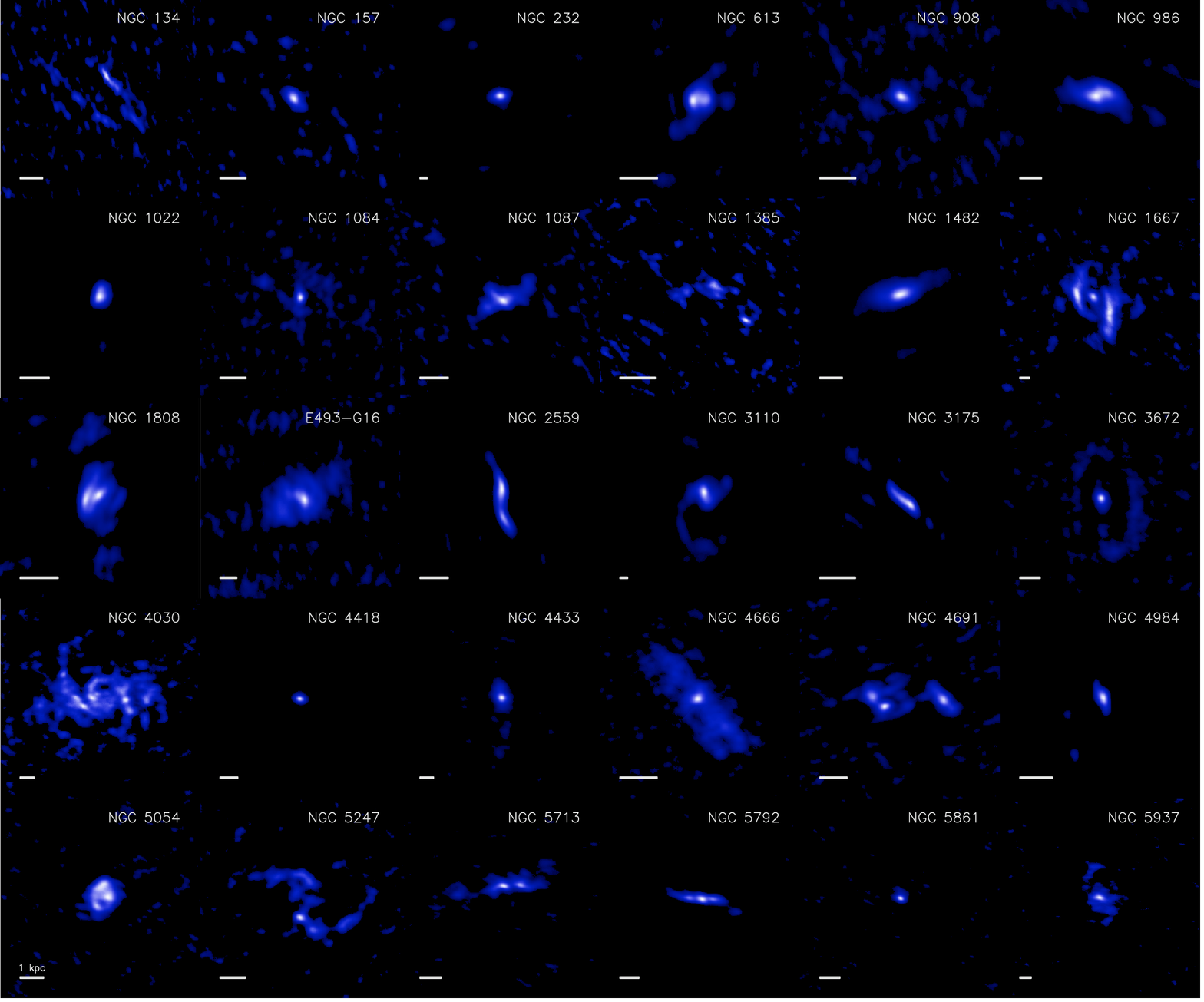}
\caption{$Up)$ DSS optical maps of the 30 galaxies in the B0DEGA subsample. The F.O.V of the images is 220\arcsec . The boxes indicate the 66\arcsec\ F.O.V of the CO(2--1) maps. The  horizontal segment on the leften side of each plot represents a 5~kpc linear scale. $Down)$ CO(2--1) distribution of the 30 galaxies in the B0DEGA subsample. The F.OV. is 66\arcsec\ , sligthly larger than the 52\arcsec\ HPBW of SMA at 230~GHz. 
The  horizontal segment on the leften side of each plot represents a 1~kpc linear scale.}
\label{fig:mom0}
\end{center}
\end{figure}

The galaxies reside in mid density regions in the local Universe, usually members of pairs, triplets or groups, thus representing the galaxies that have suffered a recent interaction.  In most cases they have peculiarities in the optical, such as asymmetries and dust lanes, and physical companions are found quite nearby in some cases.
 Most of the galaxies are in the morphological type range Sa -- Sc, although the distribution is mainly dominated (70\%) by Sb -- Sc galaxies (see optical DSS images in Fig.~\ref{fig:mom0}). A total of 72\% of the galaxies of the here presented subsample possess a bar feature.

\section{Preliminary Results: CO(2--1) morphologies and concentration}
\label{subsec:singledish}

We have produced channel maps, integrated maps (moment 0), velocity fields (moment 1), CO(2--1) spectra,  and
azimuthally averaged radial distribution profiles for each of the galaxies.  All
these data products will be publicly available at \verb!http://b0dega.iaa.es!.

 We show the CO(2--1) distribution maps for our initial subsample of 30 galaxies in Fig.~\ref{fig:mom0}.
It is remarkable that we detect molecular gas in the central regions in all of the sources.
These maps reveal a wide variety of molecular gas morphologies: extended disks (i.e. NGC~1482, NGC~4666), ring-like structures (i.e. NGC~134, 157, 1084), large scale arms (i.e. NGC~3110, NGC~4030), nuclear arms (i.e. NGC~613, NGC~986, NGC~1087, NGC~2559, NGC~5247), as well as elongated nuclear barred structures (i.e. NGC~1385, NGC~4691 and NGC~5247).

It is remarkable that the molecular gas distribution peak at the center for 86\% of the galaxies in this subsample. 
 When we restrict to galaxies for which the linear scale is better than 300~pc/\arcsec, in total $N$ = 24 galaxies, we find that these galaxies have a characteristic Gaussian width for its circumnuclear components of $<$ 1~kpc, and in average of about 0.5~kpc. An exception is NGC~4030, which has a width of about 2~kpc.
The galaxies that do not peak at the center show ring structures or nuclear bars with nuclear spiral arms emerging from it.
These galaxies are  NGC~134, NGC~1385, 4691 and 5247. NGC~134 shows a wide ring feature at 1.5~kpc. NGC~1385, NGC~4691 and NGC~5247 have quite asymmetric and elongated structures, probably nuclear bars, and with spiral arms emerging from them (NGC~4691 and NGC~5247).

\begin{figure}[]
\begin{center}
\includegraphics[width=10cm]{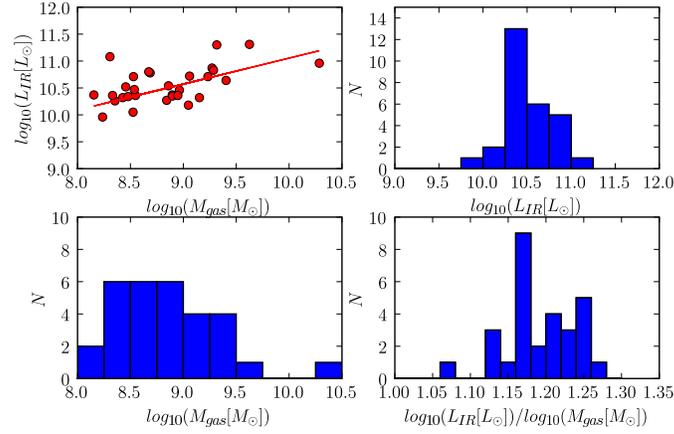}
\label{fig:sl}
\caption{$a)$ $M_{gas}$ vs $L_{IR}$ in decimal logarithmic scale, as well as a line showing the fit to the data points. The slope is 1.21 $\pm$ 0.05; $b)$ $L_{IR}$  distribution; $c)$ $M_{gas}$ distribution; $d)$ $M_{gas}$ / $L_{IR}$ distribution.  }
\end{center}
\end{figure}

We plot in Fig.~\ref{fig:sl} the global star formation (SF) law represented as the 8 -- 1000$\mu m$ FIR luminosity versus the total gas mass $M_{gas}$ calculated as 1.36 $\times$ M$_{H_{2}}$, where the 1.36 factor is to take into account species other than hydrogen. The scatter is not large if we take into account that we might have missed extended CO(2--1) emission in our interferometric observations, and suggests that most of the SF is taking place in the inner 1\arcmin, where most of the molecular gas is.

With this preliminar study, we do not find any correlation between gas concentration and existence of AGN. Actually only $\sim$ 10\% of the galaxies in this subsample are confirmed AGNs (NGC~1667, NGC~4418 and probably NGC~613), although note that the classification may not be complete.
Large starbursts are apparent, especially for barred galaxies such as NGC~1808 and NGC~613 as well as in galaxies with the largest optical perturbations, such as NGC~1022. Most of our galaxies ($\sim$ 70\%) have been classified as starburst and/or nuclear HII.

 \section{Discussion}

\begin{figure}[h]
\begin{center}
\includegraphics[width=10cm]{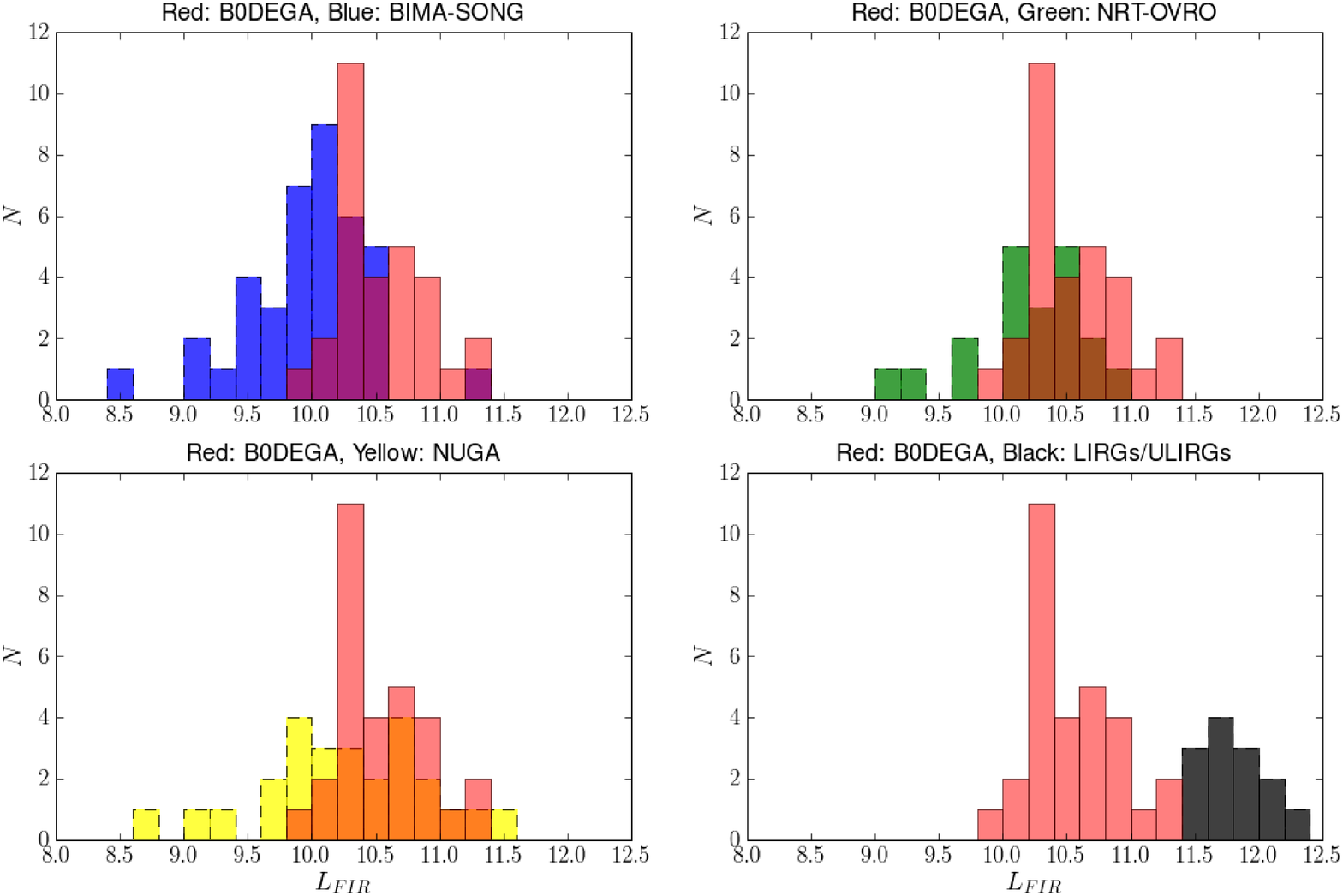}
\caption{\footnotesize  Comparison of the L$_{IR}$ distribution of our sample with that of other studies in the literature: BIMA-SONG (blue histogram), NUGA (yellow) and LIRG/ULIRGs (black).}
\label{fig:lfir-comparison}
\end{center}
\end{figure}

 Note that most of the galaxies with CO interferometric maps in the literature are IR-bright galaxies too. In particular, although the selection criteria of the BIMA-SONG, NUGA, NRT-OVRO samples are different, $\sim$ 90\% of the galaxies there fulfil our IR-flux criterion.   The distribution of IR luminosity in the B0DEGA subsample spans a range from 10.0 to 11.3~L$_\odot$, with an average of 10.6~L$_\odot$ (see Fig.~\ref{fig:lfir-comparison}). The $L_{IR}$ distribution in our sample presents in average larger values with respect to the previously mentioned samples of nearby galaxies. This effect is specially important with the BIMA-SONG sample. However, it shows low values when compared with ultra/luminous IR objects (LIRGs/ULIRGs) with $L_{IR}$ $>$ 11~L$\odot$, which are usually strongly interacting or merger systems located at larger distances. This indicates that our sample represents galaxies whose properties are between normal galaxies and mergers. On the other hand, the stellar content distribution as traced by L$_{B}$ for our galaxies is not significantly different to that of other samples, suggesting that our objects have an enhancement of SF rate when compared with similarly massive and nearby galaxies.
 
The rate of galaxies with molecular gas peaking at the center seems to be larger (by $\sim$ 45\%) in our subsample than in BIMA-SONG galaxies, which in principle is not likely to be due just by the use of different transitions of CO (J=2--1 and J=1--0, respectively). Thus, the larger concentration rate of molecular gas in our sample may mean that it is partly caused by recent interaction events, which probably trigger bar formation and other mechanisms that drive a large amount of gas to the center. Note that the rate of bars in BIMA-SONG and our subsample is similar, 67\% versus 73\%. However, in our sample the central regions might be in an evolutionary stage where the gas driven toward the center has not been fully consumed. Despite the central concentrations, most of the galaxies have not been classified to host an AGN. On the other hand, starburst and nuclear HII are apparent in many of the galaxies.

\begin{acknowledgement}
We thank the SMA  staff members who made the observations reported here possible. DE was supported by a Marie Curie International Fellowship within the 6$\rm ^{th}$ European Community Framework Programme (MOIF-CT-2006-40298).
\end{acknowledgement}

\end{document}